\documentclass{ws-procs975x65}

\begin{document}

\title{NEUTRON STAR CRUST BEYOND THE WIGNER-SEITZ APPROXIMATION}

\author{N. CHAMEL}

\address{Institute of Astronomy and Astrophysics, Universit\'e Libre de Bruxelles,\\
Brussels, Belgium\\
E-mail: nchamel@ulb.ac.be\\
www.astro.ulb.ac.be}

\begin{abstract}
For more than three decades, the inner crust of neutron stars, formed of a
solid lattice of nuclear clusters coexisting with a gas of electrons and neutrons, 
has been traditionally studied in the Wigner-Seitz approximation. The validity of 
this approximation is discussed in the general framework of the band theory of solids, 
which has been recently applied to the nuclear context. Using this novel approach, it 
is shown that the unbound neutrons move in the crust as if their mass was increased. 

\end{abstract}


\bodymatter

\section{Introduction}\label{intro}

The interpretation of many observational neutron star phenomena 
 is intimately related to the properties of neutron star crusts~\cite{haensel-06}: 
pulsar glitches, giant flares and toroidal oscillations in 
Soft Gamma Repeaters, X-ray bursts and superbursts, initial 
cooling in quasi-persistent Soft X-ray Transients, precession,  
gravitational wave emission. Besides the ejection of neutron star crust matter 
into the interstellar medium has been invoked as a promising site 
for r-process nucleosynthesis, which still remains 
one of the major mysteries of nuclear astrophysics~\cite{arnould-07}. 

The outer layers of the crust, at densities above $\sim 10^6$ g.cm$^{-3}$, are 
formed of a solid Coulomb lattice of fully ionized atomic nuclei. Below $\sim 6\times 10^{10}$ g.cm$^{-3}$, the 
structure of the crust is well established and is completely determined by the experimental atomic 
masses~\cite{ruster-06}. It is found that nuclei become increasingly neutron rich with increasing depth. At
densities above $\sim 6\times 10^{10}$ g.cm$^{-3}$, nuclei are so exotic that experimental data are  
lacking. Nevertheless, the masses of these nuclei could be determined in the near future 
by improving experimental techniques. At density $\sim 4.10^{11}$ g.cm$^{-3}$, calculations predict 
that neutrons ``drip'' out of nuclei forming a neutron liquid. Such an environment is 
unique and cannot be studied in the laboratory. We will refer to the nuclei in the inner crust 
as nuclear ``clusters'' since they do not exist in vacuum. The inner crust extends from 
$\sim 4.10^{11}$ g.cm$^{-3}$ up to $\sim$ $10^{14}$ g.cm$^{-3}$, at which point the clusters dissolve into a uniform mixture 
of electrons and nucleons. Near the crust-core interface, the clusters might adopt unusual shapes 
such as slabs, rods or more complicated structures~\cite{pethick-95, haensel-06}.

\begin{figure}[t]
\begin{center}
\psfig{file=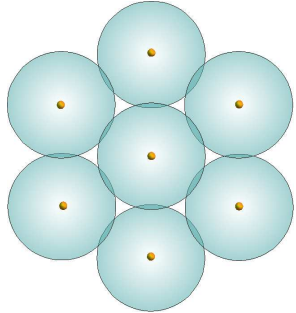,width=1.5in}
\end{center}
\caption{In the Wigner-Seitz approximation, the crust (represented here as a two dimensional hexagonal lattice) 
is divided into independent identical spheres, centered around each site of the lattice. The radius of 
the sphere is chosen so that the volume of the sphere is equal to $1/n_{\rm N}$, where $n_{\rm N}$ is 
the density of lattice sites (clusters).}
\label{fig.WS-approx}
\end{figure}

The structure of the inner crust has been studied using different approximations and nuclear models. 
The current state-of-the-art is represented by self-consistent quantum calculations 
which were pioneered by Negele\&Vautherin in 1973~\cite{nv-73}. In these calculations, the crust 
is decomposed into an arrangement of identical spheres centered around each cluster as 
illustrated in Fig.~\ref{fig.WS-approx}. Each sphere can thus be seen as one big ``nucleus'' so that 
the usual techniques of nuclear physics can be directly applied to study neutron star crust. 
However, it has been recently found that this approximation, proposed a long time ago by Wigner-Seitz~\cite{ws-33} 
in solid state physics, leads to unreliable predictions of the structure and composition of the crust especially 
in the bottom layers~\cite{baldo-06}. Besides this approach does not allow the study of transport properties which are essential for 
interpreting observations. A more elaborate treatment of neutron star crust beyond the Wigner-Seitz 
approximation is therefore required~\cite{chamel-07}. 

\section{Band theory of solids}\label{band_theory}

It is usually assumed that neutron stars are composed of cold catalyzed matter 
(for a discussion of deviations from this idealized situation, see for instance 
Ref.~\refcite{haensel-06}). Besides, it is rather well-established that the ground state 
of dense matter (at zero temperature) below saturation density possesses the 
symmetry of a perfect crystal. 

A crystal lattice can be partitioned into identical primitive cells, each of which 
contains exactly one lattice site. The specification of the primitive cell is not unique. 
A particularly useful choice is the Wigner-Seitz or Voronoi cell defined by the set of points 
that are closer to a given lattice site than to any other. This cell reflects the local 
symmetry of the crystal. It follows from the translational 
symmetry that a primitive cell contains all the information about the system. The single particle states 
of any species $q$ in a periodic system are characterized by a wave vector $\pmb{k}$. 
If the wavefunction $\varphi_{\pmb{k}}^{(q)}(\pmb{r})$ is known inside one cell, 
the wavefunction in any other cell can be deduced from the Floquet-Bloch theorem~\cite{grosso-00}
\begin{equation}
\label{eq.Bloch-theorem}
\varphi_{\pmb{k}}^{(q)}(\pmb{r}+\pmb{T}) = e^{ {\rm i} \pmb{k} \cdot \pmb{T} } \varphi_{\pmb{k}}^{(q)}(\pmb{r}) \, ,
\end{equation}
where $\pmb{T}$ is the corresponding lattice translation vector (which translates the initial 
cell to the other). Apart from the wave vector $\pmb{k}$, the single particle states are labelled by 
a discrete index $\alpha$ (principal quantum number) so that the single particle energy spectrum 
consists in a series of ``bands'' or sheets in $\pmb{k}$-space. This band index accounts for the local 
rotational symmetry of the lattice.

It can be shown that the single particle states (therefore the single particle energies) 
are periodic in $\pmb{k}$-space
\begin{equation}
\label{eq.k-period}
\varphi^{(q)}_{\pmb{k}+\pmb{K}}(\pmb{r})=\varphi^{(q)}_{\pmb{k}}(\pmb{r}) \, ,
\end{equation}
where the reciprocal vectors $\pmb{K}$ are defined by 
\begin{equation}
\pmb{K}\cdot\pmb{T}=2\pi N\, ,
\end{equation}
$N$ being any positive or negative integer. The set of all possible reciprocal vectors define a 
reciprocal lattice in $\pmb{k}$-space. Equation~(\ref{eq.k-period}) entails that only the wave vectors 
$\pmb{k}$ lying inside the so-called first Brillouin zone (i.e. Wigner-Seitz cell of the reciprocal 
lattice) are relevant. The first Brillouin zone can be further divided into irreducible domains 
by considering the rotational symmetry of the lattice. 

The inner crust of neutron stars is constituted of neutrons, protons and electrons. Unlike the 
situation in ordinary solids (ordinary meaning under terrestrial conditions), the electronic properties 
in neutron star crust are very simple. The matter density is so high that the Coulomb energy of the electrons 
is negligible compared to their kinetic energy. The electrons can thus be treated as a degenerate relativistic 
Fermi gas~\cite{pethick-95}. Since all the protons are bound inside nuclear clusters (except possibly in the 
bottom layers of the crust), the proton single particle states are almost independent of $\pmb{k}$ and to a good 
approximation can thus be described only by the discrete principal quantum number $\alpha$ like in isolated nuclei. In contrast, 
the effects of the nuclear lattice on the neutrons (i.e. dependence of the states on both $\pmb{k}$ \emph{and} $\alpha$) 
have to be taken into account because some neutrons are
unbound. In the following, we will consider the interface between the outer and the inner parts of neutron 
star crust where unbound neutrons appear. At densities below $\sim 10^{12}$ g.cm$^{-3}$, these unbound 
neutrons are not expected to be superfluid~\cite{monrozeau-07}. We will therefore treat the nucleons in the 
Hartree-Fock approximation with the effective Skyrme nucleon-nucleon interaction~\cite{bender-03}. Let us however point out that 
the above considerations are very general and apply to any approximation of the many-body problem. Indeed the Bloch 
states form a complete set of single particle states into which the many-body wave function can be expanded 
(for instance, the application of band theory including pairing correlations has been discussed in Ref.~\refcite{cch-05}). 

The single particle wave functions are obtained by solving
the self-consistent equations ($q=n,p$ for neutrons and protons
respectively) inside the Wigner-Seitz cell
\begin{equation}
\label{eq.HF1} 
h^{(q)}_0 \varphi^{(q)}_{\alpha\pmb{k}}(\pmb{r}) = 
\varepsilon^{(q)}_{\alpha\pmb{k}}
\, \varphi^{(q)}_{\alpha\pmb{k}}(\pmb{r})
\end{equation}
where the single particle Hamiltonian is defined by 
\begin{equation}
\label{eq.h0}
h^{(q)}_0  \equiv -\pmb{\nabla}\cdot \frac{\hbar^2}{2 m^{\oplus}_q(\pmb{r})} 
\pmb{\nabla} + U_q(\pmb{r}) -{\rm i} \pmb{W_q}(\pmb{r})\cdot 
\pmb{\nabla}\times\pmb{\sigma} \, .
\end{equation}
The effective masses $m^{\oplus}_q(\pmb{r})$, mean fields $U_q(\pmb{r})$ and
spin-orbit terms $\pmb{W_q}(\pmb{r})$ depend on the occupied single
particle wave functions. 

It follows from the Floquet-Bloch theorem~(\ref{eq.Bloch-theorem}), that the wavefunctions 
can be expressed as
\begin{equation}
\label{eq.Bloch.waves}
\varphi_{\alpha\pmb{k}}^{(q)}(\pmb{r}) = e^{ {\rm i} \pmb{k} \cdot \pmb{r} } u_{\alpha\pmb{k}}^{(q)}(\pmb{r}) \, ,
\end{equation}
where $u_{\alpha\pmb{k}}^{(q)}(\pmb{r})$ has the full periodicity of the lattice, 
$u_{\alpha\pmb{k}}^{(q)}(\pmb{r}+\pmb{T})=u_{\alpha\pmb{k}}^{(q)}(\pmb{r})$. Consequently, Eqs~(\ref{eq.HF1}) 
can be equivalently written as
\begin{equation}
\label{eq.HF2}
(h^{(q)}_0 + h^{(q)}_{\pmb{k}} ) u^{(q)}_{\alpha\pmb{k}} (\pmb{r}) = 
\varepsilon^{(q)}_{\alpha\pmb{k}}\,  u^{(q)}_{\alpha\pmb{k}}(\pmb{r})
\end{equation}
where the $\pmb{k}$-dependent Hamiltonian $h^{(q)}_{\pmb{k}}$ is defined by 
\begin{equation}
\label{eq.hk}
h^{(q)}_{\pmb{k}} \equiv \frac{\hbar^2 k^2}{2 m^{\oplus}_q(\pmb{r})} + 
\pmb{v_q}\cdot\hbar\pmb{k} \, ,
\end{equation}
and the velocity operator $\pmb{v_q}$ is defined by the commutator
\begin{equation}
\label{velocity_def}
\pmb{v_q} \equiv  \frac{1}{{\rm i} \hbar}[\pmb{r}, h_0^{(q)}]  \, .
\end{equation}
The boundary conditions are completely determined by the assumed crystal symmetry. 
According to the Floquet-Bloch theorem~(\ref{eq.Bloch-theorem}), the wavefunctions 
$\varphi^{(q)}_{\alpha\pmb{k}}(\pmb{r})$ and $\varphi^{(q)}_{\alpha\pmb{k}}(\pmb{r}+\pmb{T})$
between two opposite faces of the Wigner-Seitz cell ($\pmb{T}$ being the corresponding lattice vector) are 
shifted by a factor $e^{{\rm i} \pmb{k}\cdot\pmb{T}}$ while 
$u_{\alpha\pmb{k}}^{(q)}(\pmb{r})=u_{\alpha\pmb{k}}^{(q)}(\pmb{r}+\pmb{T})$.
Each cell is electrically neutral which means that it contains
as many protons as electrons. Equations are solved for a given baryon density, assuming
that the matter is in $\beta$-equilibrium. 

\section{Validity of the Wigner-Seitz approximation}\label{WS}

The three dimensional partial differential Eqs~(\ref{eq.HF1}) or (\ref{eq.HF2}) have to be solved for 
\emph{each} wave vector $\pmb{k}$ inside one irreducible domain of the first Brillouin Zone\footnote{The 
Hartree-Fock equations have to be solved together with Poisson's equation for determining the Coulomb part 
of the proton mean field potential.}. Such calculations are computationally very expensive (numerical methods
that are applicable to neutron star crust have been discussed by Chamel~\cite{chamel-05,chamel-06}). Since the work 
of Negele\&Vautherin~\cite{nv-73}, the usual approach has been to apply the Wigner-Seitz approximation~\cite{ws-33}. The complicated 
cell is replaced by a sphere of equal volume as shown in Fig.~\ref{fig.WS-approx}. In this approximation, the 
dependence on the crystal structure is therefore completely lost. Besides only the states with 
$\pmb{k}=0$ are considered or in other words the Hamiltonian $h^{(q)}_{\pmb{k}}$ is neglected. Eqs~(\ref{eq.HF1}) 
and (\ref{eq.HF2}) are then similar and reduce to a set or ordinary differential equations. The price to be paid
for such a simplification is that this approximation does not allow the study of transport properties 
(which depend on the $\pmb{k}$-dependence of the states). As pointed out by Bonche\&Vautherin~\cite{bonche-81}, 
two types of Dirichlet-Neumann mixed boundary conditions are physically plausible yielding 
a more or less constant neutron density outside the cluster: either the wave function or its 
radial derivative vanishes at the cell edge, depending on its parity. 

In recent calculations~\cite{magierski-02,newton-06,gogelein-07}, the Wigner-Seitz cell has been replaced by 
a cube with periodic boundary conditions. While such calculations allow for possible deformations of the 
nuclear clusters, the lattice periodicity is still not properly taken into account since the $\pmb{k}$-dependence 
of the states is ignored. Besides the Wigner-Seitz cell is only cubic for a simple cubic lattice and it is very 
unlikely that the equilibrium structure of the crust is of this type (the structure of the crust is expected 
to be a body centered cubic lattice~\cite{haensel-06}). It is therefore not clear whether these calculations, which require much more
computational time than those carried out in the spherical approximation, are more realistic. This point should
be clarified in future work by a detailed comparison with the full band theory. 

It should be remarked that the states obtained in the Wigner-Seitz approximation do not coincide with those obtained 
in the full band theory at $\pmb{k}=0$~\cite{chamel-07}. The reason is that in the Wigner-Seitz approximation 
the shape of the exact cell is approximated by a sphere. 
By neglecting the $\pmb{k}$-dependence of the states, the Wigner-Seitz approximation overestimates the neutron 
shell effects and leads to unphysical fluctuations of the neutron density, as discussed in details by 
Chamel {\it et al.}~\cite{chamel-07}. Negele\&Vautherin~\cite{nv-73} proposed to average the neutron 
density in the vicinity of the cell edge in order to remove these fluctuations. However it is not 
{\it a priori} guaranteed that such {\it ad hoc} procedure did remove all the spurious contributions to 
the total energy. This issue is particularly important since the total energy difference between different 
configurations may be very small. It has thus been found that the dependence of the equilibrium crust structure 
on the choice of boundary conditions, increases with density~\cite{bonche-81,baldo-06}. In particular, the 
Wigner-Seitz approximation becomes very unreliable in the bottom layers of the crust where the clusters nearly touch. 

While the Wigner-Seitz approximation is reasonable at not too high densities for determining the equilibrium crust structure, 
the full band theory is indispensable for studying transport properties. 

\section{Band theory and transport properties}\label{transport}

\begin{figure}[t]
\begin{center}
\psfig{file=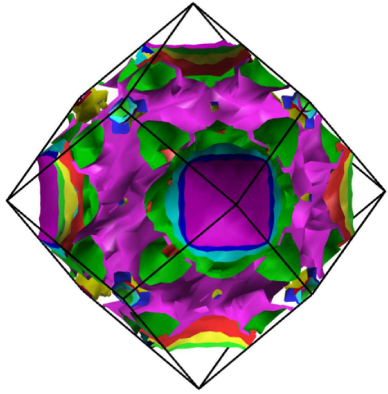,width=2.5in}
\end{center}
\caption{Neutron Fermi surface in the inner crust of neutron star at average mass density $\rho=7\times 10^{11}$ g.cm$^{-3}$. 
The crust is composed of a body centered cubic lattice of zirconium-like clusters ($Z=40$) with $N=160$ neutrons per 
lattice site ($70$ neutrons are unbound). The single particle states have been calculated by solving 
the Hartree-Fock equations with the Skyrme force SLy4 and by applying Bloch boundary conditions~\cite{chamel-07}. The 
Fermi surface is shown inside the first Brillouin zone. The different colors correspond to different bands.}
\label{fig.FS}
\end{figure}

At low temperature the transport properties of the neutron liquid in neutron star crust are determined 
by the shape of the Fermi surface, defined by the locus of points in $\pmb{k}$-space such that 
$\varepsilon_{\alpha\pmb{k}}=\varepsilon_{\rm F}$ (we drop the superscript $n$ since we only consider neutrons in 
this section). In general, the Fermi surface is composed of several sheets corresponding to the different bands 
that intersect the Fermi level. An example of neutron Fermi surface is shown on Fig.~\ref{fig.FS}. 

The motion of the unbound neutrons is affected by the presence of the nuclear clusters. In particular, they can 
be Bragg-reflected by the crystal lattice. As a result of the momentum
transfer with the lattice, a neutron in a state characterized by a wave vector $\pmb{k}$ and a band $\alpha$, move in the crystal 
as if its mass was replaced by an effective mass given by (considering cubic crystals)
\begin{equation}
\label{eq.effmass}
m_n^\star\vert_{\alpha\pmb{k}} = 3\hbar^2 \left( \Delta_{\pmb{k}} \varepsilon_{\alpha\pmb{k}}\right)^{-1}  \, ,
\end{equation}
where $\Delta_{\pmb{k}}$ denotes the Laplacian operator in $\pmb{k}$-space. This concept of effective mass 
has been very useful in interpreting neutron diffraction experiments~\cite{zeilinger-86}. In the context of neutron star crust, 
the number of unbound neutrons can be very large and it is therefore more appropriate to introduce a ``macroscopic'' 
effective mass $m_n^\star$ averaged over all occupied single particle states
\begin{equation}
\label{eq.maceffmass}
m_n^\star = n_{\rm f}/{\cal K} \, , \hskip0.5cm {\cal K}=\sum_\alpha \int_{\rm F} \frac{{\rm d}^3\pmb{k}}{(2\pi)^3}\, \left(m_n^\star\vert_{\alpha\pmb{k}}\right)^{-1} \, ,
\end{equation}
where $n_{\rm f}$ is the number density of free neutrons and the integral is performed over all occupied states 
``inside'' the Fermi surface\footnote{The occupied states are those for which the energy $\varepsilon_{\alpha\pmb{k}}$
is lower than the Fermi energy $\varepsilon_{\rm F}$.}. 
Let us remark that the mobility coefficient $\cal K$ can be equivalently written as 
\begin{equation}
{\cal K}=\frac{1}{3 (2\pi)^3\hbar^2}\sum_\alpha\int_{\rm F} \pmb{\nabla}_{\pmb{k}}\varepsilon_{\alpha\pmb{k}} \cdot \pmb{{\rm d}S}\, ,
\end{equation}
showing that the effective mass depends on the shape of the Fermi surface. 
This effective mass $m_n^\star$ has been found to be very large compared to the bare mass~\cite{chamel-05,chamel-06}. For instance, the effective mass 
corresponding to the Fermi surface shown in Fig.~\ref{fig.FS} is about $m_n^\star\simeq 4.3\, m_n$. On the contrary, this effective mass
is slightly lower than the bare mass in the liquid core owing to the absence of clusters~\cite{chamelhaensel-06} (the effectice mass $m_n^\star$ still differ from $m_n$ due to neutron-neutron and neutron-proton interactions). 
With the definition~(\ref{eq.maceffmass}), it can be shown that in the crust frame the momentum $\pmb{p_{\rm f}}$ of 
the neutron liquid is simply given by $\pmb{p_{\rm f}}=m_n^\star\, \pmb{v_{\rm f}}$, where $\pmb{v_{\rm f}}$ is the neutron velocity~\cite{cchI,cchII}. 
This result is quite remarkable. It implies that in an \emph{arbitrary} frame, the momentum and the velocity of the neutron liquid 
are not aligned! Indeed, if $\pmb{v_{\rm c}}$ is the velocity of the crust, applying the Galilean transformation leads to
\begin{equation}
\label{eq.entrainment}
\pmb{p_{\rm f}}=m_n^\star\, \pmb{v_{\rm f}} + (m_n - m_n^\star) \pmb{v_{\rm c}}\, .
\end{equation}
These so-called entrainment effects are very important for the dynamics of the neutron liquid in the crust. For instance, it has been suggested 
that for sufficiently large effective mass $m_n^\star \gg m_n$, a Kelvin-Helmholtz instability could occur and might explain the origin of pulsar
glitches~\cite{andersson-04}.

\section{Conclusion}

Since the work of Negele\&Vautherin~\cite{nv-73}, the Wigner-Seitz approximation has been widely 
applied to study neutron star crust. Nevertheless, the necessity to go beyond has become clear 
in the last few years. It has thus been shown that the results of the Wigner-Seitz 
approximation become less and less reliable with increasing density~\cite{baldo-06}. Besides
this approach, which decomposes the crust into a set of independent spherical cells, 
does not allow the study of transport properties. A more realistic description 
of the crust requires the application of the band theory of solids~\cite{chamel-07}. By
taking consistently into account both the nuclear clusters which form a solid lattice
and the neutron liquid, this theory provides a unified scheme for studying the properties
of neutron star crust. We have shown in this framework that the unbound neutrons move in the crust 
as if they had an effective mass much larger than the bare mass. 
The analogy between neutron star crust and periodic systems in 
condensed matter suggests that some layers of the crust might have properties similar to those
of band gap materials like semiconductors or photonic crystals for instance. The existence of 
such ``neutronic'' crystals in neutron star crusts opens a new interdisciplinary field of research, 
 at the crossroad between nuclear and solid state physics.

\section*{Acknowledgments}

N.C. gratefully acknowledges financial support from a Marie Curie Intra
European fellowship (contract number MEIF-CT-2005-024660). 

\bibliographystyle{ws-procs975x65}
\bibliography{chamel}

\begin{thebibliography}{10}

\bibitem{haensel-06}
P.~{Hansel}, A.~Y. {Potekhin} and D.~G. {Yakovlev}, {\em Neutron Stars 1 :
  Equation of State and Structure} (Springer, 2006).

\bibitem{arnould-07}
M.~{Arnould}, S.~{Goriely} and K.~{Takahashi}, {\em Phys. Rep.} {\bf 450}, 97
  (2007).

\bibitem{ruster-06}
S.~B. {R{\"u}ster}, M.~{Hempel} and J.~{Schaffner-Bielich}, {\em Phys. Rev. C}
  {\bf 73}, 035804 (2006).

\bibitem{pethick-95}
C.~J. {Pethick} and D.~G. {Ravenhall}, {\em Ann. Rev. Nucl. Part. Sci.} {\bf
  45}, 429 (1995).

\bibitem{nv-73}
J.~W. {Negele} and D.~{Vautherin}, {\em Nucl. Phys. A} {\bf 207}, 298 (1973).

\bibitem{ws-33}
E.~{Wigner} and F.~{Seitz}, {\em Phys. Rev.} {\bf 43}, 804 (1933).

\bibitem{baldo-06}
M.~{Baldo}, E.~E. {Saperstein} and S.~V. {Tolokonnikov}, {\em Nucl. Phys. A}
  {\bf 775}, 235 (2006).

\bibitem{chamel-07}
N.~{Chamel}, S.~{Naimi}, E.~{Khan} and J.~{Margueron}, {\em Phys. Rev. C} {\bf
  75}, 055806 (2007).

\bibitem{grosso-00}
G.~{Grosso} and G.~P. {Parravicini}, {\em Solid State Physics} (Elsevier,
  2000).

\bibitem{monrozeau-07}
C.~{Monrozeau}, J.~{Margueron} and N.~{Sandulescu}, {\em Phys. Rev. C} {\bf
  75}, 065807 (2007).

\bibitem{bender-03}
M.~{Bender}, P.~{Heenen} and P.~{Reinhard}, {\em Rev. Mod. Phys.} {\bf 75}, 121
  (2003).

\bibitem{cch-05}
B.~{Carter}, N.~{Chamel} and P.~{Haensel}, {\em Nucl. Phys. A} {\bf 759}, 441
  (2005).

\bibitem{chamel-05}
N.~{Chamel}, {\em Nucl. Phys. A} {\bf 747}, 109 (2005).

\bibitem{chamel-06}
N.~{Chamel}, {\em Nucl. Phys. A} {\bf 773}, 263 (2006).

\bibitem{bonche-81}
P.~{Bonche} and D.~{Vautherin}, {\em Nucl. Phys. A} {\bf 372}, 496 (1981).

\bibitem{magierski-02}
P.~{Magierski} and P.-H. {Heenen}, {\em Phys. Rev. C} {\bf 65}, 045804 (2002).

\bibitem{newton-06}
W.~G. {Newton}, J.~R. {Stone} and A.~{Mezzacappa}, {\em J. Phys. Conf. Series}
  {\bf 46}, 408 (2006).

\bibitem{gogelein-07}
P.~{G{\"o}gelein} and H.~{M{\"u}ther}, {\em Phys. Rev. C} {\bf 76}, 024312
  (2007).

\bibitem{zeilinger-86}
A.~{Zeilinger}, C.~G. {Shull}, M.~A. {Horne} and K.~D. {Finkelstein}, {\em
  Phys. Rev. Lett.} {\bf 57}, 3089 (1986).

\bibitem{chamelhaensel-06}
N.~{Chamel} and P.~{Haensel}, {\em Phys. Rev. C} {\bf 73}, 045802 (2006).

\bibitem{cchI}
B.~{Carter}, N.~{Chamel} and P.~{Haensel}, {\em Int. Mod. Phys. D} {\bf 15},
  777 (2006).

\bibitem{cchII}
B.~{Carter}, N.~{Chamel} and P.~{Haensel}, {\em Nucl. Phys. A} {\bf 748}, 675
  (2005).

\bibitem{andersson-04}
N.~{Andersson}, G.~L. {Comer} and R.~{Prix}, {\em MNRAS} {\bf 354}, 101 (2004).

\end{thebibliography}

\end{document}